\documentstyle[prl,preprint,aps]{revtex}
 
\begin{document}
\draft

\widetext

\title{Thermodynamics of the incommensurate state in $Rb_{2}WO_{4}$:\\
on the Lifshitz point in $A^{\prime }A^{\prime \prime }BX_{4}$  compounds}

\author{I. Luk'yanchuk}
\address {L.D.Landau Institute for Theoretical Physics, 117940, Moscow,
Russia\\
and Departamento de F\' \i sica, Universidade Federal de Minas Gerais\\
Caixa Postal 702, 30161-970, Belo Horizonte, Minas Gerais, Brazil}

\author{A. J\'orio}
\address{Departamento de F\' \i sica, Universidade Federal de Minas Gerais\\
Caixa Postal 702, 30161-970, Belo Horizonte, Minas Gerais, Brazil}

\author{P. Saint-Gr\'egoire}
\address{L.M.M.I., Universit\'e de Toulon et du Var, BP 132, 83957 La Garde
c\'edex, France}

\date{\today }
\maketitle

\widetext
\begin{abstract}

We consider the evolution   of the  phase transition from the parent
 hexagonal
phase $P6_{3}/mmc$ to the orthorhombic phase $Pmcn$ that occurs in several
compounds of $A^{\prime }A^{\prime \prime }BX_{4}$  family as a function of
the hcp 
lattice parameter $c/a$. For compounds
of $K_{2}SO_{4}$ type with $c/a$ larger than the threshold value 1.26
the direct first-order transition $Pmcn-P6_{3}/mmc$ is characterized by the
large entropy jump $\sim R\ln 2$. For compounds $Rb_{2}WO_{4}$,
$K_{2}MoO_{4}$, $K_{2}WO_{4}$ with $c/a<1.26$ this transition occurs via an
intermediate
incommensurate $(Inc)$ phase. DSC measurements were performed in
$Rb_{2}WO_{4}$ to characterize the thermodynamics of the
 $Pmcn-Inc-P6_{3}/mmc$
transitions. It was found that both transitions are again of the first order
with entropy jumps $0.2\cdot R\ln 2$ and $0.3\cdot R\ln 2$.
Therefore, at $c/a \sim 1.26$ the $A^{\prime }A^{\prime \prime }BX_{4}$  
compounds reveal an unusual Lifshitz
point where three first order transition lines
meet.  We propose the coupling of crystal elasticity with  $BX_{4}$
 tetrahedra
orientation as a possible source of the  transitions
discontinuity.

\end{abstract}
\pacs{PACS: 64.70.Rh, 64.60.Cn}

\newpage  

\setlength{\parindent}{5pt} \leftskip -10pt \rightskip 10pt

\section{Introduction}

The orientational ordering of $BX_{4}$ tetrahedra drives a rich sequence of
structural phases in ionic $A^{\prime }A^{\prime \prime }BX_{4}$ compounds
of $K_{2}SO_{4}$ type. In the present communication we are interested in the
nature of the phase transition from the parent high-symmetry phase $%
P6_{3}/mmc$ (like $\alpha $-$K_{2}SO_{4}$) to the orthorhombic phase $Pmcn$
(like $\beta $-$K_{2}SO_{4}$) that occurs at high temperatures ($T\sim
600-800K$) either directly: 
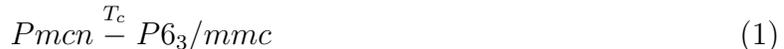
\begin{equation}
Pmcn\stackrel{T_{c}}{-}P6_{3}/mmc  \label{Dir}
\end{equation}
(e.g. in $K_{2}SO_{4}$, $Rb_{2}SeO_{4}$, $K_{2}SeO_{4}$) or via an
intermediate $1q-$incommensurate ($Inc$) phase:

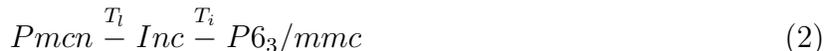
\begin{equation}
Pmcn\stackrel{T_{l}}{-}Inc\stackrel{T_{i}}{-}P6_{3}/mmc  \label{Inc}
\end{equation}
as in molybdates and tungstates $Rb_{2}WO_{4}$, $K_{2}MoO_{4}$,
$K_{2}WO_{4}$. The last one has the modulation vector ${\bf
q= }(0,q_{b},0)\,$ that can be
alternatively directed in two other equivalent directions of the $120^{0}$
star of the hexagonal Brillouin zone. All the transitions are of the
order-disorder type and are characterized by the vertical (up/down)
orientations of $BX_{4}$ tetrahedra. Other, low temperature transitions in
 $%
A^{\prime }A^{\prime \prime }BX_{4}$ compounds that are related with the
planar
orientation of tetrahedra are beyond our consideration (for details, see
Refs. \cite{Kur,KurRev,L}).

  From the
viewpoint of the Landau theory of phase transitions, only the
lock-in  $Pmcn-Inc$ transition should be of the first order. The
 $Pmcn-P6_{3}/mmc$ transition should be of the second order since $Pmcn$ is
 a
subgroup of $P6_{3}/mmc$ and in the Landau functional neither third order
nor Lifshits terms are present. The transition $Inc-P6_{3}/mmc$ should also
be of the second order as a transition to the incommensurate phase of the
type II. 
The recently proposed hcp Ising model \cite{L} correctly describes
the high-temperature phase diagram of $A^{\prime }A^{\prime \prime }BX_{4}$
compounds. In this model the  
$Pmcn-P6_{3}/mmc$ and $Inc-P6_{3}/mmc$ transitions are of the second order.

The experimental properties of $\ $the $Pmcn-P6_{3}/mmc$ transitions in
various compounds of $A^{\prime }A^{\prime \prime }BX_{4}$ family are
collected from Refs.\cite{G,M,F,S,Rao,Lop,W} in Table I as function of the
geometrical factor $c/a$ of their hcp structure. As was shown in our
previous study \cite{L} this is the unique parameter that drives the actual
phase sequence: for $c/a>1.26$ the transition is direct, whereas for $%
c/a<1.26$ the sequence (\ref{Inc}) takes place.

In disagreement with the theoretical prediction, the direct
 $Pmcn-P6_{3}/mmc$
transition is of the first order with a large jump of the molar entropy
($\sim R\ln 2$) \cite{S,Lop} and of the lattice constants ($\sim 2\%$)
\cite{M,F}. 
The Incommensurate phase in $Rb_{2}WO_{4}$, $K_{2}MoO_{4}$, $K_{2}WO_{4}$
 was
relatively poorly studied because of the high hygroscopic nature of these
compounds 
\cite{W,M1}. It is known that the $Inc-P6_{3}/mmc$ transition reveals a 
substantial discontinuity of the lattice parameter ($\sim 0.2-0.7\%$) \cite
{W}. To characterize the thermodynamics of the $Pmcn-Inc-P6_{3}/mmc$
transition we performed  Differential Scanning Calorimetry (DSC)
measurements in $Rb_{2}WO_{4}$ that are reported in Sec. II. It was
found that both transitions are of the first order with entropy jumps of $%
0.2R\ln 2$ and $0.3R\ln 2$. The $\ Inc-P6_{3}/mmc$ transition is a rare
example of incommensurate transition that occurs discontinuously.

At $c/a\sim 1.26$ the critical temperatures $T_{l}$, $T_{i}$, $T_{c}$
coincide 
and the $A^{\prime }A^{\prime \prime }BX_{4}$ compounds seem to reveal a 
triple Lifshitz point, that was found previously only in few experimental
systems (for a review  see Ref. \cite{UFN}). The particular property
of this Lifshitz point in $A^{\prime }A^{\prime \prime }BX_{4}$ compounds
is that all the incoming transition lines are of the first order. This
possibility was theoretically studied in Ref. \cite{NaNo} where the
discontinuities were modeled by the negative forth-order terms in the Landau
functional. To our knowledge this is also the unique example of Lifshitz
point in a system where the modulation vector can be directed in more than
one equivalent direction.

The main question raised by these systems is why strong discontinuities at
$Pmcn-P6_{3}/mmc$ and $Inc-P6_{3}/mmc$ transitions appear. Note first that
they cannot be
ascribed to fluctuation effects that were widely studied last decades in
relation with transitions to the modulated phases \cite
{Braz,Kats,Horn,Muk,Bar}. In such a case the first order character is
attributed to the lack of a stable fixed point as in $BaMnF_4$ \cite{StGreg}
 and the discontinuity is expected to be
small due to the smallness of the critical region. We propose that the
observed discontinuities are caused by the coupling of the order parameter
with elasticity of the crystal that is known \cite{Domb,Sal,Lar} to be able
to change the order of transition. Introducing in Sec. III the corresponding
coupling to the mean-field treatment of the hcp Ising model \cite{L} and
comparing the results with the measured jumps of the lattice constant and
molar entropy we demonstrate that this coupling can be
responsible for the transitions discontinuity.

\section{Experiment}

DSC experiments were performed in $Rb_{2}WO_{4}$ crystals to characterize
the thermodynamics of the $Pmcn-Inc-P6_{3}/mmc$ transitions. Due to the very
high hygroscopic nature of the material, powder samples were prepared
in a special camera in a dry nitrogen atmosphere. DSC experiments have been
performed using a Mettler-TA3000 equipment, between room temperature and $%
820K$. The heating/cooling rate was $5K/min$. DSC thermograms of the
investigated sample show the presence of two reversible enthalpic anomalies
at about $T_{l}= 660K$ and $T_{i}= 746K$, the lock-in and the
 incommensurate
phase transitions, respectively (see Fig. I). The measured molar entropy
jumps are $\Delta S_{Tl}= 1.4J/K\cdot mol$ and $\Delta S_{Ti}= 1.8J/K\cdot
mol$ (approximately $80\%$ of $\Delta S_{Ti}$ are taken from the $\delta -$%
peak of DSC anomaly at $T_{i}$ and other $20\%$ from the residual specific
heat decrease in the temperature interval of about $8K$ below $T_{i}$, see
Fig. I). These values are given in the Table I in units of $R\ln 2$. An
hysteresis of $12.5K$ was observed for the $Pmcn-Inc$ transition, ($%
T_{l}= 664\pm 0.5K$ for heating and $651.5\pm 0.5K$ for cooling). In
contrast, the $Inc-P6_{3}/mmc$ transition reveals no hysteresis within the
error bar of $\pm 0.5K$. This is consistent with extremely small hysteresis
of $1K$ observed for $Pmcn-P6_{3}/mmc$ transition in $K_{2}SO_{4}$
 \cite{Rao}.

\section{Discussion}

The high-temperature order-disorder transitions in $A^{\prime }A^{\prime
\prime }BX_{4}$ compounds are described by the in-site averages of the
vertical orientation of $BX_{4}$ tetrahedra, $\sigma _{i}= <S_{i}>$ 
where the pseudo-spin $S_i$ is equal to $\pm 1$ for the up/down tetrahedra
orientations \cite{L}.
The variables $\sigma_i$ are equal to zero in the disordered
high-temperature phase $P6_{3}/mmc$. In the low-temperature phase $Pmcn$,
they take the equal amplitudes $\sigma _{i}= \pm \sigma $ and alternate
according to $Pmcn$ symmetry.
In the incommensurate phase a modulation 
$\sigma _{i}= \sigma _{q}(e^{i{\bf qr}_{i}}+e^{-i{\bf qr}_{i}})= 2\sigma
_{q}\cos {\bf qr}_{i}$ occurs. The
absolute values of $\sigma _{i}$, and hence of amplitudes $\sigma $ and $%
2\sigma _{q}$ (that define the corresponding order parameters), are
smaller than one; the less they are, the more $BX_{4}$ tetrahedra are
disordered. Because of the discontinuity of the $Pmcn-P6_{3}/mmc$ and $%
Inc-P6_{3}/mmc$ transitions the amplitudes $\sigma $ and $2\sigma _{q}$
have nonvanishing values below the critical temperatures $T_{c}$ and
$T_{i}$. We estimate $%
\sigma $ and $2\sigma _{q}$ in the ordered states from the entropy jump at
the transition:

\begin{equation}
\frac{\Delta S}{R}= <\frac{1}{2}((1+\sigma _{i})\ln (1+\sigma
_{i})+(1-\sigma _{i})\ln
(1-\sigma _{i}))>_{i}  \label{entropy}
\end{equation}
that is measured experimentally (see Table I). For $K_{2}SeO_{4}$ and $%
K_{2}SO_{4}$, the inequality $\Delta S>R\ln 2$ holds, which means that in
 the
low temperature phase the $BX_{4}$ tetrahedra are perfectly ordered ($\sigma
\sim 1$) and, possibly, other degrees of freedom are involved in the
transition. Taking $\sigma _{i}$ in the incommensurate phase of
 $Rb_{2}WO_{4}
$ as $2\sigma _{q}\cos {\bf qr}_{i}$, from $\Delta S= 0.3R\ln 2$ we get $%
2\sigma _{q}\sim 0.8$ that again demonstrates the high degree of tetrahedra
ordering.

In the mean-field approach of the hcp Ising model \cite{L}, the phase
transitions from $P6_{3}/mmc$ to $Pmcn$ and to $Inc$ phases were found to be
continuous and the free energy (per molecule) was expanded over the
small values of parameters $\sigma $ and $2\sigma _{q}$ as:

\begin{equation}
f_{com}= \frac{k}{2}(T-T_{c})\sigma ^{2}+\frac{kT}{12}\sigma ^{4}
\label{fcom}
\end{equation}
for $Pmcn-P6_{3}/mmc$ transition, and as:

\begin{equation}
f_{inc}= \frac{k}{4}(T-T_{i})(2\sigma _{q})^{2}+\frac{kT}{32}(2\sigma
_{q})^{4}  \label{finc}
\end{equation}
for $Inc-P6_{3}/mmc$ transition.

The critical temperatures $T_{c}$, $T_{i}$ are functions of interaction
parameters $J_{ij}$. They coincide at the Lifshitz point and are correlated
with
the geometrical factor $c/a$ as follows: $T_{c}<T_{i}$ when $a/c<1.26$ and
 $%
T_{c}>T_{i}$ when $a/c>1.26$.

To account for the discontinuity of the transitions we propose that
coupling of
tetrahedra orientation with crystal elasticity is responsible for this
phenomenon. Our further consideration is analogous to compressible Ising
model, proposed by Domb \cite{Domb}. For estimation purpose we consider here
only the coupling with the strain $e_{3}$ along the hexagonal axis and omit
other elastic degrees of freedom. Account of those is difficult because of
the absence of experimental data, but it can only improve our estimations.

The elastic contribution to the free energy is written as: 
\begin{equation}
f_{el}= \gamma \sigma ^{2}e_{3}+\frac{1}{2}V_{mol}C_{33}e_{3}^{2}
\label{felcom}
\end{equation}
where $\gamma \sigma ^{2}e_{3}$ is the coupling of the order parameter with
elastic strain and $V_{mol}C_{33}e_{3}^{2}$ is the proper elastic energy of
the crystal (per unit volume $V_{mol}$ of the molecule). After minimization
we get the strain in the $Pmcn$ phase: $e_{3}= \Delta c/c= -\gamma \sigma
^{2}/V_{mol}C_{33}$. Substituting it back to (\ref{felcom}) we find that
coupling with elastic strain renormalizes the quartic term in (\ref{fcom})
and the total free energy is written as:

\begin{eqnarray}
f_{com}+f_{el} &= &\frac{k}{2}(T-T_{c})\sigma ^{2}  \label{ftotcom} \\
&&+(kT/12-\gamma ^{2}/2C_{33}V_{mol})\sigma ^{4}  \nonumber
\end{eqnarray}

The quartic term becomes negative when the elastic contribution $\gamma
^{2}/2C_{33}V_{mol}$ exceeds the Ising thermal energy $kT/12$. The
transition then is of the first order and the amplitude of $Pmcn$ order
parameter is stabilized by the higher order terms.

We estimate the value of the coupling constant $\gamma $ from the relation
 $%
\gamma $ =  $-\Delta c/c\cdot V_{mol}C_{33}/\sigma ^{2}$ that, in
 $K_{2}SO_{4}
$ with $C_{33}= 55\cdot 10^{9}N/m^{2}$ \cite{C33}, $\sigma ^{2}\sim 1,$ $%
V_{mol}\sim 120\AA ^{3}$ \cite{M} and $\Delta c/c= 0.025$ gives $\gamma
 \sim
-1.7\cdot 10^{-19}J$. Then, the elastic contribution $\gamma
^{2}/2C_{33}V_{mol}\sim 2.1\cdot 10^{-21}J$ is indeed larger than $%
kT_{c}/12= 10^{-21}J$ that justifies the role of the elastic degrees of
freedom in the discontinuity of $Pmcn-P6_{3}/mmc$ transition.

Consider now the $Inc-P6_{3}/mmc$ transition. Assuming that the elastic
coupling is given by $\gamma <\sigma^2_{i}> e_{3}$= $\frac{1}{2}\gamma
(2\sigma _{q})^{2}e_{3}$, we come to the effective
functional

\begin{eqnarray}
f_{inc}+f_{el} &= &\frac{k}{4}(T-T_{i})(2\sigma _{q})^{2}  \label{ftotinc}
 \\
&&+(kT/32-\gamma ^{2}/8C_{33}V_{mol})(2\sigma _{q})^{4}  \nonumber
\end{eqnarray}

We perform the estimation of the quartic coefficient for $Rb_{2}WO_{4}$ in
an analogous way as the precedent, taking $2\sigma _{q}\sim 0.8$ and $%
V_{mol}\sim 142\AA ^{3}$ \cite{W}. Since  the elastic constant
 $%
C_{33}$ is not available we assume it has the same value as in $K_{2}SO_{4}$. The jump
of the lattice parameter $\Delta c/c$ is assumed to be of the same order of 
 $0.007$ as in $K_{2}MoO_{4}$. The calculation gives: $\gamma = -2\Delta
c/c\cdot V_{mol}C_{33}/(2\sigma _{q})^{2}\sim -2\cdot 10^{-19}J$ and $\gamma
^{2}/8C_{33}V_{mol}\sim 13\cdot 10^{-22}J$ that again is larger than the 
 bare
forth-order coefficient $kT_{i}/32= 3\cdot 10^{-22}J$.

To conclude, we suggest that the Lifshitz point occurs in $A^{\prime
}A^{\prime \prime }BX_{4}$ compounds at $c/a\sim 1.26$ where three
first-order transition lines meet together. One can expect to achieve this
point experimentally, either by preparation of solid solution
$Rb_2W_xMo_{1-x}O_4$ or by submitting $K_{2}SeO_{4}$ or $Tl_{2}SeO_{4}$ 
(with $c/a= 1.27$ and $1.26$) to an uniaxial pressure along $c$. Analyzing
experimental
data we demonstrate that the coupling of the order parameter with crystal
elasticity can be responsible for the discontinuity of transition. 

We stress another peculiar feature of the $Pmcn-P6_{3}/mmc$ and $%
Inc-P6_{3}/mmc$ transitions. Despite of the strong entropy jump ($\sim R\ln 
 2
$), they have a very low hysteresis (less than $1K$) that cannot be
explained on the basis of the available models \cite{Domb,Sal}. 

It is interesting to note that the $Pmcn-P6_{3}/mmc$ transition occurs also 
 in
another compound of $A^{\prime }A^{\prime \prime }BX_{4}$ family - 
 $KLiSO_{4}
$ that has a large ratio $c/a= 1.69$. 
Unlike other cases, this transition is either
of the second or of the weak first order with  the entropy jump less then $%
0.1R\ln 2$ \cite{B} and with no visible jump  of the lattice constants \cite
{Bob}. Then, it is quite probable, that the order of transition changes from
the first to the second one and that the  $Pmcn-P6_{3}/mmc$ transition line
reveals a tricritical point when  $c/a$ increases. More systematic
experiments however are needed to verify this hypothesis.

We are grateful to M. A. Pimenta, R. L. Moreira, W. Selke, A. S. Chaves, 
F. C. de S\'{a} Barreto, J. A. Plascak for helpful discussions and to A. M.
Moreira for technical assistance. The work of I. L. was supported by the
Brazilian Agency Fundacao de Amparo a Pesquisa em Minas Gerais (FAPEMIG) and
by Russian Foundation of Fundamental Investigations (RFFI), Grant No.
960218431a The work of A. J. was supported by the
Brazilian Agency Funda\c c\~ao Coordena\c c\~ao de Aperfei\c coamento 
de Pessoal de N\' \i vel Superior (CAPES).

\begin{figure}[t]
  
\caption{DSC thermograms of $Rb_{2}WO_{4}$ in the region of 
$Pmcn\stackrel{T_{l}}{-}Inc\stackrel{T_{i}}{-}P6_{3}/mmc$ phase 
transitions on cooling and on heating.} 
 
\label{fig1}
\end{figure}

\begin{table}[tbp]
\caption{The critical temperatures $T_{l}$, $T_{i}$ for $Pmcn$ - $Inc$ - $%
P6_{3}/mmc$ transitions and $T_{c}$, for $Pmcn$ - $P6_{3}/mmc$ transitions,
the molar entropy jumps and the lattice parameter jumps as functions of
lattice parameter $c/a$. The entropy jumps in $Rb_{2}WO_{4}$ were measured
in the present study.} 
 
\begin{tabular}{cccccccc}
& $c/a$ & $T_{l},K$ & $T_{i},K$ & $(\frac{\Delta S}{R\ln 2})_{Tl}$ & $(\frac{%
\Delta S}{R\ln 2})_{Ti} $ & $(\frac{\Delta c}{c})_{Tl}$ & $(\frac{\Delta c}{c%
})_{Ti}$ \\ 
&  & \multicolumn{2}{c}{$T_{c},K$} & \multicolumn{2}{c}{$(\frac{\Delta S}{%
R\ln 2})_{Tc} $} & \multicolumn{2}{c}{$(\frac{\Delta c}{c})_{Tc}$} \\ \hline
$K_{2}WO_{4}$ & $1.24$ & $643$ & $707$ &  &  & $<0.2\%$ & $0.2\%$ \\ 
$K_{2}MoO_{4}$ & $1.24$ & $593$ & $733$ &  &  & $<0.2\%$ & $0.7\%$ \\ 
$Rb_{2}WO_{4}$ & $1.25$ & $660$ & $746$ & $0.2$ & $0.3$ &  &  \\ \hline
$Tl_{2}SeO_{4}$ & $1.26$ & \multicolumn{2}{c}{$660$} & \multicolumn{2}{c}{}
& \multicolumn{2}{c}{} \\ 
$K_{2}SeO_{4}$ & $1.27$ & \multicolumn{2}{c}{$745$} & \multicolumn{2}{c}{$%
1.3 $} & \multicolumn{2}{c}{} \\ 
$Rb_{2}MoO_{4}$ & $1.27$ & \multicolumn{2}{c}{$775$} & \multicolumn{2}{c}{}
& \multicolumn{2}{c}{} \\ 
$Rb_{2}SeO_{4}$ & $1.29$ & \multicolumn{2}{c}{$825$} & \multicolumn{2}{c}{}
& \multicolumn{2}{c}{} \\ 
$Cs_{2}SeO_{4}$ & $1.29$ & \multicolumn{2}{c}{$860$} & \multicolumn{2}{c}{}
& \multicolumn{2}{c}{} \\ 
$K_{2}SO_{4}$ & $1.29$ & \multicolumn{2}{c}{$860$} & \multicolumn{2}{c}{$2.1$%
} & \multicolumn{2}{c}{$2.5\%$} \\ 
$Tl_{2}SO_{4}$ & $1.30$ & \multicolumn{2}{c}{$773$} & \multicolumn{2}{c}{} & 
\multicolumn{2}{c}{$<1.3\%$}
\end{tabular}

\end{table}



\begin{references}
\bibitem{Kur}  M. Kurzy\'{n}ski and M. Halawa, Phys. Rev., {\bf B34}, 4846
(1986).

\bibitem{KurRev}  M. Kurzy\'{n}ski, Act. Phys. Pol., {\bf B6}, 1101 (1995)

\bibitem{L}  I.Luk'yanchuk, A. J\'orio and M. A. Pimenta, Phys.Rev. {\bf
B57}, 5086
(1998)

\bibitem{G}  G. Gatow, Acta Cryst., {\bf 15}, 419 (1962)

\bibitem{M}  A.J. Majumdar and R. Roy, J. Phys. Chem. {\bf 69}, 1684 (1965)

\bibitem{F}  G. Pannetier and M. Gaultier, Bull. Soc. Chim. Fr., 1069 (1966)

\bibitem{S}  C. H. Shomate and B. F. Naylor, J. Am. Chem. Soc., {\bf 67}, 72
(1945)

\bibitem{Rao}  C. N. R. Rao and K. J. Rao {\it Phase Transitions in Solids},
(McGraw-Hill Inc., 1978)

\bibitem{Lop}  A. Lopez Echarri, M. J. Tello and P. Gili, Sol. St. Com., 
{\bf 36}, 1021 (1980)

\bibitem{W}  J. Warczewski, Phase Transitions, {\bf 1} , 131 (1979)

\bibitem{M1}  F. Tunistra and A. J. van den Berg, Phase Transitions, {\bf 3}
, 275 (1983) and refs. therin

\bibitem{UFN}  Yu. M. Vysochanski\u {\i } and V. Yu. Slivka, Usp. Fiz. Nauk,=
 
{\bf 162}, 139 (1992) [Sov. Phys. Usp. {\bf 35}, 123 (1992)]

\bibitem{NaNo}  S. L. Qui, Mitra Dutta, H. Z. Cummins, J. P. Wicked and S.
M. Shapiro, Phys. Rev. {\bf B34}, 7901 (1986)

\bibitem{Braz}  S. A. Brazovsky, I. E. Dzyaloshinsky and A. R. Muratov, Zh.
Eksp. Teor. Fiz. {\bf 75}, 1140 (1987) [Sov. Phys. JETP, {\bf 48}, 573
(1987)]

\bibitem{Kats}  E. I. Kats, V. V. Lebedev and A. R. Muratov, Phys. Reports
 
{\bf 228}, 1 (1993)

\bibitem{Horn}  R. M. Hornreich, M. Luban and S. Schtrikman, Phys. Rev.
Lett., {\bf 35}, 1678 (1975)

\bibitem{Muk}  D. Mukamel and M. Luban, Phys. Rev., {\bf 18}, 3631 (1978)

\bibitem{Bar}  C. Barbosa, Phys. Rev. {\bf B42}, 6363 (1990)

\bibitem{StGreg} P. Saint-Gr\'egoire, W. Kleemann, F. J. Schafer and J. Moret, J. Physique {\bf 49}, 463 (1988)

\bibitem{Domb}  C. Domb, J. Chem. Phys. {\bf 25}, 783 (1956)

\bibitem{Sal}  S. R. Salinas, J. Phys. C: Solid State Phys., {\bf 7}, 241
(1974) and refs. therin.

\bibitem{Lar}  A. I. Larkin and S. A. Pikin, Zh. Eksp. Teor. Fiz. {\bf 56},
1664 (1969) [Sov. Phys.-JETP, {\bf 29}, 891 (1969)]

\bibitem{C33}  LANDOLT-B\"{O}RNSTEIN, New  Series, Vol. III/18, {\it %
Elastic, Piezoelectric and related constants of crystals, }Edited by
K.-H.Hellwege and A.H.Hellwege, (Springer, 1984)

\bibitem{B}  T. Breczewski, P. Piskunowicz and G. Jaroma-Weiland, Acta Phys.
Polonica, {\bf A66} (1984)

\bibitem{Bob}  A. Righi and R. L. Moreira, private communication.
\end{references}
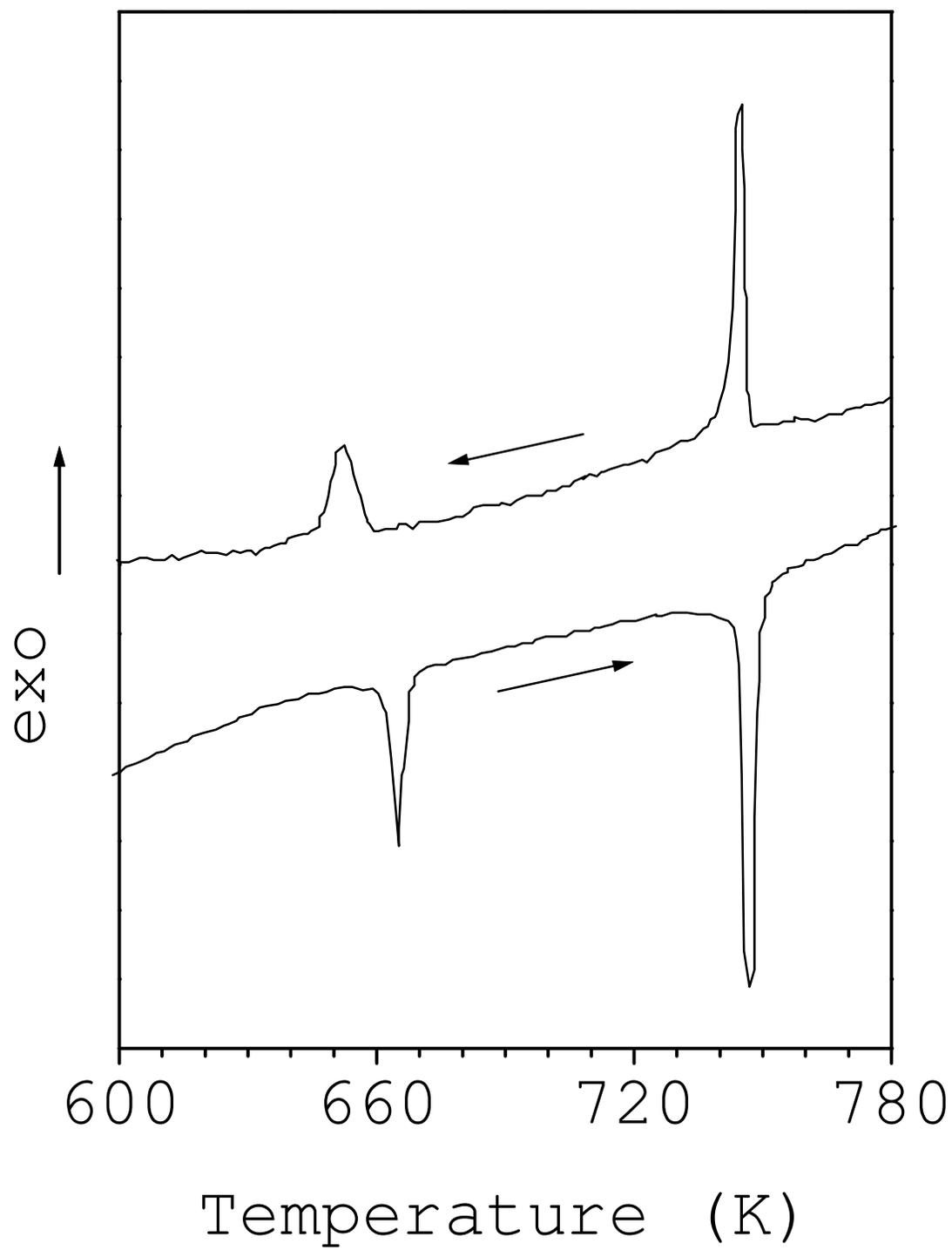
\end{document}